\documentclass[letterpaper,10pt]{article}
\usepackage{osameet2}

\usepackage{amsmath,amssymb}
\usepackage{booktabs}
\usepackage[pdftex,colorlinks=true,bookmarks=false,citecolor=blue,urlcolor=blue]{hyperref} %pdflatex

\begin{document}

\title{Defragmentation-as-a-Service (DaaS): \\How beneficial is it?}

\author{Sandeep Kumar Singh, Wolfgang Bziuk, and Admela Jukan}
\address{Technische Universit\"at Carolo-Wilhelmina zu Braunschweig, Germany}
\email{\{sandeep.singh, w.bziuk, a.jukan\}@tu-bs.de}

\begin{abstract}
We analyze Defragmentation-as-a-Service (DaaS) in elastic optical networks and show that the positive effect of defragmentation depends on the rate at which it is performed, load (or, call arrival rates) and the available resources. \end{abstract}

%\ocis{000.0000, 999.9999.}

\section{Introduction}
\par The concept of continuity is natural when it comes to service, otherwise its counterpart, i.e. discontinuity (or fragmentation)  can hurt the system performance. Fragmentation is an age old problem in storage devices, and recently it also was shown to affect the network performance in elastic optical networks (EON). In EON, spectrum can be sliced to a smaller units (e.g., 6.25 GHz) to suit the diverse applications. While serving heterogeneous application demands for bandwidth, setting up and termination of connection requests can cause spectrum fragmentation of an elastic optical link (EOL), which results in higher blocking probability. To alleviate this problem, network defragmentation, - similar to the disk defragmentation, can be used to reconfigure some of the connections such that largest possible free resource block are created for the future requests. However, reconfiguration of a connection causes  disruption of its traffic for the duration of retuning time of transceivers   and  reconfiguration time of the optical switches. One can avoid reconfiguration time of switches  by making the connection over new frequency slots (FS) before breaking it from its existing path, however migration (retuning) will still cause some interruption.

\par In this paper, we coin the term Defragmentation-as-a-Service (DaaS) in elastic optical networks, and analyze the  positive effect of defragmentation (DF) process as a function of the rate at which it is performed, load (or, call arrival rates) and the available resources. We also analyze the impact of DF rate on the blocking probability for different spectrum allocation (SA) policies, such as first-fit (FF) and random-fit (RF) \cite{yu2014exact,beyranvand2014analytical,Rosa2015}.
To understand the effect of defragmentation, we model an EOL using a multi-class continuous-time Markov chain (CTMC), and include DaaS states in addition to the normal service\footnote{We refer to \emph{normal service} for the case of service which does not employ defragmentation} states. We have the following assumptions for our model: i) the DF process is called when a request is blocked due to the fragmentation of a link; ii) all existing requests are interrupted during DF period, iii) all incoming requests are blocked during the DF period, and iv) system returns to normality, i.e. services of existing requests are resumed when DF is completed.
Although assumptions, particularly (ii) and (iv), are stronger, traffic interruption of some (but not all) existing connections does happen during reconfiguration in the real system. Note that the DaaS states are blocking states for all types of connections, hence if DF is performed at lower rate then we would experience an increase in overall blocking probability (BP). Using our model, we show that DF reduces the overall call blocking probability only if: a) it is performed at the higher rates as compare to the normal service rate of the calls, b) load on system is moderate (i.e., call arrival rates are not high), and c) resources (bandwidth) are sufficient.
\vspace{-2mm}
\section{Proposed Model: Defragmentation-As-a-Service}
\begin{figure}[t]
 \centering
\includegraphics[width= 1\textwidth]{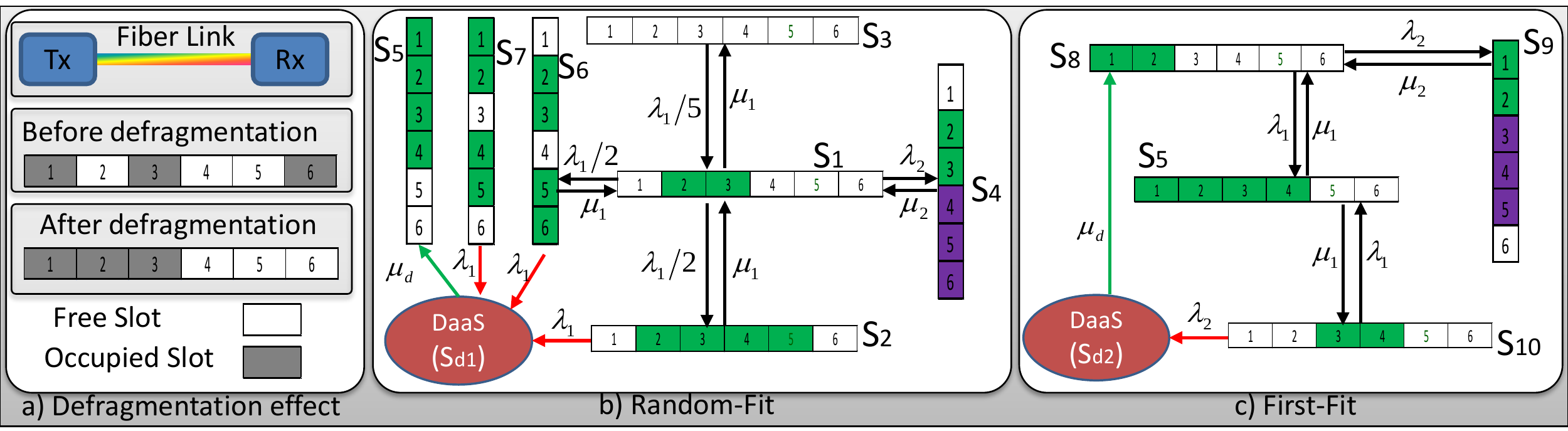}
\vspace{-8mm}
  \caption{DaaS: its effect, and state transitions leading to defragmentation for RF and FF SA policies.}
\label{fig:system}
\vspace{-2mm}
\end{figure}
\vspace{-1mm}
We adopt a multi-class CTMC model to analyze the blocking probability of EOL.
Here, we assume that a DaaS process is triggered when a request is blocked due to fragmentation of an EOL, and not due to the unavailability of resources. To model the reconfiguration of the occupied FS, we introduce DaaS states. We assume that both inter-arrival  and call holding times of class-k requests are independent and  exponentially distributed with average rates $\lambda_k$ and  $\mu_k$, respectively. In Fig. \ref{fig:system}(a), let us consider an example of an EOL link with $C=6$ FS to illustrate the idea.  In Fig. \ref{fig:system}(b) and (c), we consider two classes of requests with demands $2$ and $3$ consecutive slots for RF and FF SA policies, respectively and show some of the transitions, due to the arrivals and departures, leading to a DaaS state. In contrast to FF, which allocates first available consecutive slots to a request, RF randomly allocates one of the possible assignments. In Fig. \ref{fig:system}(b), under the RF method, consider a case where a new request (say $R_1$) of class $k=1$ arrives in state $S_1$ with demand $d_1=2$. There are $a(S_1,k)=2$ different ways to occupy the $3$ free slots of state $S_1$, thus transitions ($S_1 \rightarrow S_2$ and $S_1 \rightarrow S_6$) occurs with rates $\lambda_1/2$.
Furthermore, when a class-1 request (two slots) arrives, blocking of the request due to fragmentation occurs in the set of states $FB(1)=\{S_2, S_6, S_7\}$. 
These blocking events will trigger the transition from states $S_i \in FB(1)$  to the DaaS state $S_{d1}$ (red arrows). During the exponentially distributed DaaS time $T_{DF}$ with mean $1/\mu_d$, arriving calls are blocked. Furthermore, services of existing connections effected by spectrum reconfiguration are interrupted during $T_{DF}$. Finally, system returns to a non-fragmented state (green transition to $S_5$), where services of the existing requests are resumed. Due to the memoryless property of the exponential distribution, the remaining call holding time is again exponentially distributed with mean $1/\mu_k$. Thus, transitions due to call terminations are modeled with rates $\mu_k$.
\vspace{-3mm}
\begin{figure}[ht!]
 \centering
\includegraphics[width=1\textwidth, height =3.8in]{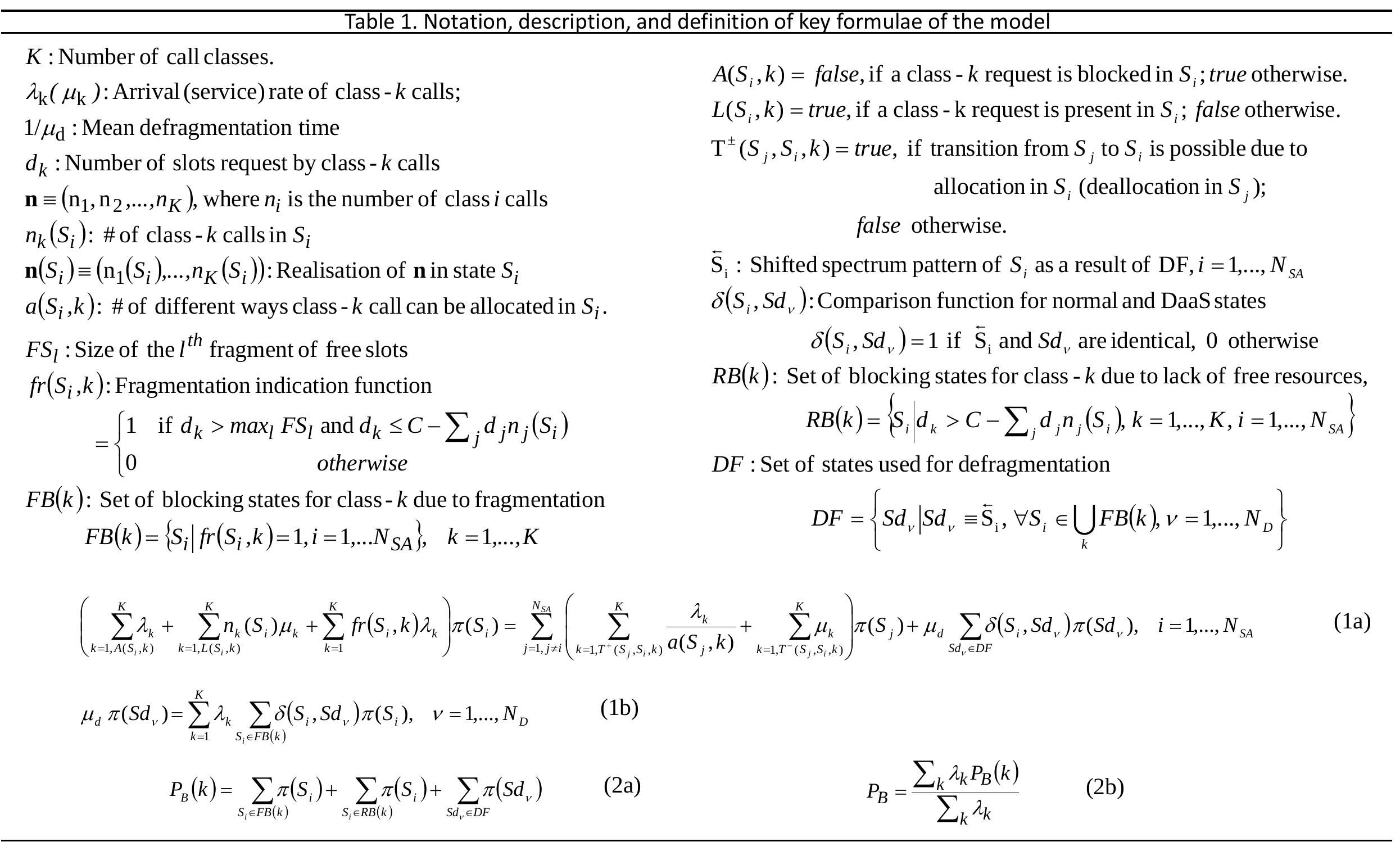}
\label{table:formulae}%
\vspace{-7mm}
\end{figure}
The global balance equations (GBE) for the RF policy are presented in Table 1. Eq.(1a) gives the GBE for normal occupancy states $S_i, i=1,...,N_{SA}$, whereas (1b) gives the GBE for DaaS states $Sd_\nu, \nu=1,...,N_D$. The stationary state probabilities  $\pi = [ \pi(S_1),...,\pi(S_{N_{SA}}), \pi(Sd_1),...,\pi(Sd_{N_D})]$ can be calculate by solving $\pi$ $Q=0$ subject to $\sum_i \pi(S_i)+\sum_\nu \pi(Sd_\nu)=1 $, where $Q$ is the transition rate matrix. For example, the GBE for state $S_1$ is given by $(\lambda_1 + \lambda_2 + \mu_1)\pi(S_1) = \lambda_1/5 \pi(S_3)+ \mu_1(\pi(S_2)+\pi(S_6)) +\mu_2\pi(S_4)$. The left (right) side characterizes the probability flow out of (into) $S_1$. The input rate $\lambda_1/a(S_3,1)=\lambda_1/5$ takes into account that 6 free slots of $S_3$ results into 5 different spectrum pattern, one of which is $S_1$. For the output flow we have $n_1(S_1)=1$, since there is only one existing class-1 call. The DaaS-state $Sd_1$ is reached from $S_i \in FB(1)=\{S_2,S_6,S_7\}$, which have different spectrum pattern, but results in same pattern after shifting, i.e. $\overleftarrow{S_i}\equiv  Sd_1, i = 2,6,7$. Thus, the function $\delta(S_i,Sd_1)=1, i=2,6,7$ assigns to the right DaaS-state, and $\delta(S_5,Sd_1)=1$ links to the final de-fragmented state. Using (1b), the GBE of state $Sd_1$ is given by $\mu_d \pi(Sd_1)=\lambda_1 (\pi(S_2)+\pi(S_6)+\pi(S_7))$. The FF policy is also modeled by Eq.(1), taking into account that some transitions are not allowed due to the assignment strategy. Particularly, we have $a(S_i,k) = 1$. A more detailed presentation of the model is omitted for sake of brevity. The blocking probability of a class-k request is given in (2a), and the overall blocking in (2b). In our model, blocking occurs due to the fragmentation ($1^{st}$ term in (2a)), resource unavailability ($2^{nd}$ term), and also in DaaS states (last term).
 
\vspace{-2mm}

\section{Numerical Results}
\begin{figure}[t]
 \centering
\includegraphics[width=1\textwidth, height=2.4in]{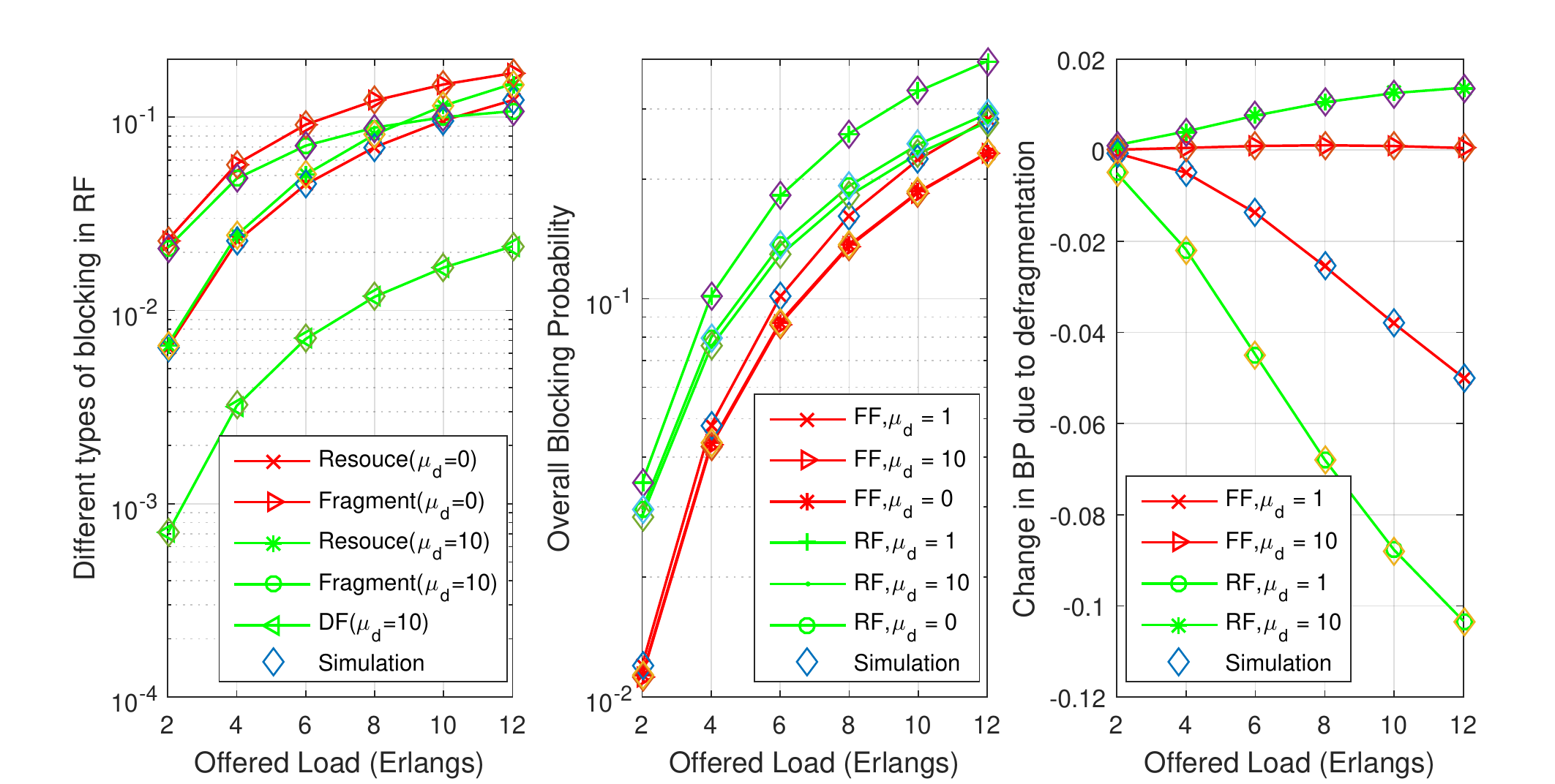}
\vspace{-8mm}
  \caption{Effect of DaaS when applied to RF and FF SA policies.}
\label{fig:defrag}
\vspace{-2mm}
\end{figure}
We show the analytical and verifying Monte-Carlo simulation results for an EOL having 20 FS, and three classes of requests with demand $d_1=4, d_2=6$ and $d_3=8$ slots. Arrival rates are uniformly distributed and mean holding time is one unit for all classes of requests. Load on the link is calculated as $\sum_{k =1 }^ K\frac{\lambda_k d_k}{\mu_k}$. DaaS rates of $\mu_d=1$ and $\mu_d=10$ are used. A regular system without DaaS is shown by $\mu_d=0$. In Fig. \ref{fig:defrag}(left), we compare the blocking parts under RF policy (Eq.(2a)) due to the effects of lack of resources, fragmentation and defragmentation. In contrast to other work, our analytical model allows us to show accurate results for the important practical range of blocking probabilities below $10\%$. As expected, without DaaS and for a suitable traffic load, the blocking is dominated by fragmentation and not by resource unavailability. For $\mu_d=10$, the DaaS  reduces the fragmentation blocking at cost of the resource blocking, which becomes dominant for higher load ($>6$ Erl.). The additional blocking due to the DaaS is insignificantly lower (about two magnitudes lower). As expected (in Fig. \ref{fig:defrag}(center)), when average defragmentation service time ($t_{DF}=1/\mu_d$) is same as the average service time of the class-k connection (i.e., $t_{Sk}=1/\mu_k$), then blocking probabilities are higher as compared to the system without DaaS in both FF and RF methods. To get the advantage of DaaS, $t_{DF}$ must be lower than the average service time $t_{Sk}$. As shown, all parts of the blocking probability reduces significantly for RF and marginally for FF methods when we increase the DaaS rate from 1 to 10, as corresponding service time decreases proportionally. We show the change in blocking probabilities in Fig. \ref{fig:defrag}(right) as the difference in blocking probabilities of with and without DaaS. First we notice, that the FF policy has overall no significant advantage of defragmentation, also for small $t_{DF}$. For RF, the achieved gain (change in BP) is positive for a smaller $t_{DF}$ (= 1/10) for the load range shown here, however, the gain decreases at higher loads due to the negative effect of DaaS. The reason is that when arrival rates are higher, DaaS states would be called very frequently, which increases blocking. 
\vspace{-2mm}

%\vspace{-1mm}
\section{Conclusion}
We model for the first time the defragmentation-as-a-service and showed the correlation among DF service rate, load and the available resources. The results showed that over exploitation of DF can hurt  the system performance. 


\vspace{-2mm}
\begin{thebibliography}{99}
\bibitem{yu2014exact} Y. Yu \emph{et al.}, ``Exact performance analytical model for spectrum allocation in flexible grid optical networks,'' \emph{Journal of Optical Fiber Technology}, vol. 20, no. 2, pp. 75--83, 2014.

\bibitem{beyranvand2014analytical} H. Beyranvand \emph{et al.}, ``An Analytical Framework for the Performance Evaluation of Node-and Network-Wise Operation Scenarios in Elastic Optical Networks," \emph{IEEE Trans. on Netw.}, vol. 62, no. 5, pp. 1621--1633, 2014.
\bibitem{Rosa2015}
Rosa, A. N. F., et al. ``Statistical analysis of blocking probability and fragmentation based on Markov modeling of elastic spectrum allocation on fiber link." \emph{Optics Communications} 354 (2015): 362-373.

\end{thebibliography}
\end{document}